\documentclass[conference]{IEEEtran}

\IEEEoverridecommandlockouts
\usepackage{cite}
\usepackage{amsmath,amssymb,amsfonts}
\usepackage{algorithmic}
\usepackage{graphicx}
\usepackage{textcomp}
\usepackage{multirow}
\usepackage{url}
\usepackage{comment}
\usepackage{ulem}
\usepackage{xcolor}
\usepackage{lipsum}
\usepackage{array}

\begin{document}
\title{Automatic Issue Classifier: A Transfer Learning Framework for Classifying Issue Reports}
 
 %1. \title{Automatic Issue Classification using Transfer Learning}

% 2. AIC: Automatic Issue classifier
 
% 3. Attention based Issue Classification
 
% 4. Multi-Label ISSUE classification using Attention based MODEL
 
% 5. 

\author{
\IEEEauthorblockA{Anas Nadeem}
\IEEEauthorblockA{\textit{ Dept. of Computer Science } \\
\textit{North Dakota State University}\\
Fargo, USA \\
anas.nadeem@ndsu.edu}
\and
\IEEEauthorblockA{Muhammad Usman Sarwar}
\IEEEauthorblockA{\textit{ Dept. of Computer Science } \\
\textit{North Dakota State University}\\
Fargo, USA \\
muhammad.sarwar@ndsu.edu}
\and
\IEEEauthorblockN{Muhammad Zubair Malik}
\IEEEauthorblockA{\textit{Dept. of Computer Science} \\
\textit{North Dakota State University}\\
Fargo, USA \\
zubair.malik@ndsu.edu}

}

\maketitle
\begin{abstract}

Issue tracking systems are used in the software industry for the facilitation of maintenance activities that keep the software robust and up to date with ever-changing industry requirements. Usually, users report issues that can be categorized into different labels such as bug reports, enhancement requests, and questions related to the software. Most of the issue tracking systems make the labelling of these issue reports optional for the issue submitter, which leads to a large number of unlabeled issue reports. In this paper, we present a state-of-the-art method to classify the issue reports into their respective categories i.e. bug, enhancement, and question. This is a challenging task because of the common use of informal language in the issue reports. Existing studies use traditional natural language processing approaches adopting key-word based features, which fail to incorporate the contextual relationship between words and therefore result in a high rate of false positives and false negatives. Moreover, previous works utilize a uni-label approach to classify the issue reports however, in reality, an issue-submitter can tag one issue report with more than one label at a time. This paper presents our approach to classify the issue reports in a multi-label setting. We use an off-the-shelf neural network called RoBERTa and fine-tune it to classify the issue reports. We validate our approach on issue reports belonging to numerous industrial projects from GitHub. We were able to achieve promising F-1 scores of 81\%, 74\%, and 80\% for bug reports, enhancements, and questions, respectively. We also develop an industry tool called Automatic Issue Classifier (AIC), which automatically assigns labels to newly reported issues on GitHub repositories with high accuracy.

\end{abstract}
\section{Introduction}
\label{sec:introduction}
Software maintenance refers to the changes made to the software in order to make it more robust, efficient, and to make it adaptable to a modified environment \cite{software1998ieee}. Issue tracking systems are one of the most critical means for the maintainers to enable rigorous yet effective software evolution tasks. Issue tracking systems allow the software developers to report, manage and keep track of the tickets \cite{KALLIS2021102598}. They are capable of being the single point of reference for nearly all the maintenance activities such as resolving reported bugs, making the project more efficient, and adding new features.

GitHub is a version control and source code management service, which empowers the users in several ways such as to host the source code, keeping track of versions, and release and maintain the software. It is the largest source code hosting platform with more than 50 million developers and more than 100 million public repositories \footnote{\url{https://venturebeat.com/2018/11/08/github-passes-100-million-repositories/}} as of the year 2018. Similar to the other conventional bug-tracking systems such as Jira\footnote{\url{https://www.atlassian.com/software/jira}}, MantisBT\footnote{\url{https://www.mantisbt.org/}}, and Zoho\footnote{\url{https://www.zoho.com/bugtracker/}}, GitHub also provides an integrated issue tracking system. Using the GitHub issue tracking system, a user can interact with the maintainers of the repository. For instance, any user who wants to report a bug, request a new feature, or wants to ask questions related to the software might make use of the issue tracking system.% to communicate with maintainers.

To create an issue report \footnote{Terms 'issue' and 'issue report' are used interchangeably throughout the paper.} in the GitHub issue tracker, users are required to provide a title and a brief description as the body. However, the labels associated with the issue are optional. Issue submitter can choose these labels from the list of default issue labels such as 'bug', 'enhancement', 'documentation', and 'question' or it can be custom defined by the issue submitter. Additionally, the issue submitter can tag the issue report with multiple labels at a time. Labelling these issues enables the project team to prioritize these issues based on the label assigned. For example, while the software is close to a release, it might be important to make the software more robust instead of adding new features. While all these categorized issue reports help the software maintainers to identify, reproduce and finally make appropriate changes in the software, it also complicates the tasks for the software maintainer due to the inconsistent nature of labels. For instance, the bug-report label can be written in multiple ways such as 'bug', 'there is a bug' etc., which makes this a challenging task. Usually, software project repositories either rely on the issue-submitter to assign a label to the issue report based on their intuition or they rely on the software development team to manually classify these issues to their respective categories. The manual classification approach works well for repositories in which issues are reported with a low frequency. However, there are numerous popular repositories on GitHub which have a very large number of issues reported every day, making it a challenging task to manually label them. Also, labelling these issues on such repositories is a non-trivial task and generally requires specific domain knowledge and a thorough analysis of the body of the issue.

Classification of GitHub issue reports has recently been of great interest as an enabling path for a more involved GitHub-based data analysis. However, the existing approaches to solve the issue report classification problem do not use advanced language modeling and rely on keyword-based approaches \cite{KALLIS2021102598,chawla,antonio,FAZAYELI2019585,fan}. Traditional, keyword-based approaches suffer from a high false-positive and a high false-negative rate. This is because they view the problem as bag-of-words and are not aware of the full language model, hence they do not capture contextual relationships among the words \cite{sarwar}. Also, the existing studies view the problem as a multi-class classification problem that assumes an issue can be only associated with a single label. However, an issue submitter can tag the issue label with multiple labels at the same time. Therefore, we have to treat the issue classification problem as multi-label classification. In multi-label classification settings, an issue report can be associated with more than one label at a time.

To address the aforementioned problems, we used a transformer-based model called RoBERTa \cite{roberta}. RoBERTa is an optimized variant of BERT, which uses improved training methodology and larger data to achieve better results than BERT\cite{devlin2018bert}. RoBERTa has the ability to consider the contextual relationship between the word while making the prediction. To train RoBERTa, we take the GitHub issue data set labeled with the 3 labels i.e. `Bug', `Enhancement', `Question'. Here, we only used issues that are tagged with the default GitHub labels. As we are essentially performing transfer learning, we need significantly fewer data samples to build an accurate model as compared to millions of data examples required to train a model from scratch \cite{sarwar}. We evaluated our approach on approximately 55,000 randomly sampled issue reports associated with the top 200 repositories across 55 popular languages. We were able to achieve promising results with an F1-Score nearing 80\% with significantly fewer training examples.

In our work, we tried to answer the following research questions:
\begin{enumerate}
    \item RQ1: Does the transformer-based approach out-perform the traditional keyword-based approaches at classifying GitHub issue reports?
    \item RQ2: Does the transformer-based model have the ability to be incorporated into an industry-level tool?

\end{enumerate}

This work has immediate benefits to the industry as it enables efficient tracking of issue reports on GitHub. First, it can help the software development team track the software maintenance process effectively. Secondly, the categorized issue classification can help in effective development documentation that can help the project team to achieve effective software maintenance. Thirdly, having accurately categorized issue reports would help in routing the issue to the right software developer, who is working on the specific module. Lastly, categorized issue reports can help the project manager to optimize the resource allocation process, if a particular module is getting a lot of bug reports that implies the need of assigning resources for the purpose of improving the associated code-base.

%We evaluate our results on actual GitHub repositories and were able to achieve an F1 score of above 80\% with lesser training required, which is a significant improvement over existing approaches. This work has immediate benefits to the industry as it enables easy tracking of issues on GitHub. Additionally, we release all the data collected, the complete source code for our GitHub application as well as the code for building the machine learning model to enable future work.

The following is a summary of our contributions:

\begin{itemize}
    \item We present a transformer-based approach for classifying the GitHub issue reports in a multi-label setting.
    \item We conducted our evaluation on GitHub issue-tracker dataset across 55 different programming languages.
    \item We develop and release a tool, Automatic Issue Classifier (AIC) on top of our transfer-based approach. AIC is capable of being deployed in industrial-scale software repositories to automatically classifies issue reports as soon as they are reported.
\end{itemize}

This paper is organized as follows: Section \ref{sec:related_work} discusses the previous relevant studies. Section \ref{sec:methodology} discusses the methodology we used for the data collection, and issue report categorization. Section \ref{sec:evaluation} discusses the evaluation of our approach. Section \ref{sec:discussion} discusses the implication of these results in the industry. Section \ref{sec:threats} discusses the threats to validate the study. Section \ref{sec:future_work} discusses the potential future avenues. Finally, section \ref{sec:conclusion} concludes the study.

%\begin{figure}
% \centerline{\includegraphics[width=0.25\textwidth]{Figures/default_labels.png}}
% \caption{A list of default GitHub labels available for a user while creating an issue report}
% \label{fig:default_github_labels}
%\end{figure}

%\begin{figure}
% \centerline{\includegraphics[width=0.5\textwidth]{Figures/complete_issue_form.png}}
% \caption{A complete overview of the issue reporting 0process along-with a list of default GitHub labels available for a user while creating an issue report}
% \label{fig:default_github_labels}
%\end{figure}
\section{Related Work}
\label{sec:related_work}

Previous studies in the domain of classification of issues on issue-trackers range from the binary classification of bugs/non-bugs to the categorization of issues using multi-class classification methods. Several research works have proposed methods ranging from using keyword-based approaches to sophisticated machine learning models. We discuss these studies and their limitations in order to lay the grounds for our research.

Antoniol et al.\cite{antonio} proposed a key-word based approach in order to distinguish bug issues from non-bug issues. They manually labeled 1,800 issues associated with repositories of Mozilla, Eclipse, and JBoss and further fed these issues to Alternating Decision Trees, Naive Bayes, and Logistic Regression-based models resulting in up to 82\% correct results. Our work proposes a multi-label method to classify these issues into three mutually exclusive labels. Additionally, we evaluate our results on a larger set of issues from a diverse set of repositories.

Kallis et al. \cite{KALLIS2021102598} presented TicketTagger, a tool to categorize GitHub issues. They used a fastText \cite{joulin2016bag} based model to classify the GitHub issue into bugs, enhancement, and questions categories. They were able to achieve an F1-score of 82\% over all three of the categories. They treated the issue classification problem as a multi-class problem where one issue can belong to a single label at a time.
%They treat these categories as mutually exclusive whereas originally one issue report can fall into more than one category.
Our work used the same labels as used by Kallis et al. \cite{KALLIS2021102598}. However, we solve this problem as a multi-label classification problem, where an issue report can have multiple labels at the same time.

%Our research is coherent with attempting to classify the issue reports into the same categories however, we take into account the non-exclusive nature of the issue-report labels. As a result, our tool in comparison is able to assign multiple labels to issue reports.

Fan et al. \cite{fan} proposed a two-stage framework for classifying the issue reports. They perform their analysis on over 252,000 issue reports from 80 popular GitHub projects. Further, they compared their results with four traditional machine learning methods including Naive Bayes, Logistic Regression, Random Forest Tree, Support Vector Machine. Our work, in contrast, leverages the power of pre-trained transformer-based models which produce state-of-the-art results without requiring a large number of training examples.

Chawla et al. \cite{chawla} proposed an automated approach to categorize the type of issue reports. They utilized a Fuzzy Logic based system to classify the issue types and evaluated their results on issue reports from HTTPClient, Jackrabbit, and Lucene. They were able to achieve an F-1 score of 0.83, 0.79, 0.84 for the three repositories, respectively. However, they evaluated their model on results on a limited number of well-maintained repositories which might be misleading.

Fazayeli et al. \cite{FAZAYELI2019585} used text-based classification to label GitHub issues. However, they only evaluated their framework for only a single repository i.e. 'git-for-windows'. Also, they treated the problem as a binary classification problem with bug and non-bug labels. However, our framework categorically attempts to assign multiple labels to a GitHub issue report. Also, we evaluated our approach on a diverse data set with projects from various programming languages.

The above-discussed studies utilized keyword-based features to categorize the issue reports, which suffer from a high rate of false positives. The primary reason of which is that they do not consider the contextual relationships between words while classification. Also, these studies treat the issue classification problem as a multi-class classification problem. However, in real-world scenarios, an issue can be associated with more than one label at a time and should be addressed as a multi-label classification problem.

Attention-based networks derived from human intuition have resulted in significant improvement in various natural language processing tasks\cite{attention_nlp}. Attention-based networks are capable of focusing on context-relevant details in a given text while ignoring irrelevant keywords. Bahdanau et al.\cite{bahdanau2016neural} were the first to introduce an attention mechanism for the purpose of machine translation, commonly referred to as additive attention. Recent advancements in NLP has led to wide adoption of self-attention based Transformers \cite{vaswani2017attention} models. Pre-trained transformers are trained on a large natural language corpus and further, they can be fine-tuned for the downstream NLP tasks such as machine translation, and text classification.

In recent work, Devlin et al.\cite{bert} introduced Bidirectional Transformers for Language Understanding (BERT) which leverages the Transformer-architecture. BERT was originally pre-trained on BooksCorpus containing 800M words and English Wikipedia containing 2,500 million words. This pre-training over large data sets enables BERT to get familiar with language vocabulary. Further fine-tuning the pre-trained model for specific tasks refers to Transfer Learning. There are several BERT variants based on architecture such as the number of input layers, hidden layers, self-attention heads, and parameters used. While BERT is a powerful framework for NLP tasks, tweaking it can substantially improve its performance. Liu et al.\cite{roberta} proposed RoBERTa which is a robust and optimized variant of BERT. They proposed several design changes such as removing the next sequence prediction objective, changing the masking pattern based on the data itself. They show that these changes along with pre-training the model for a longer duration and in larger batches results in substantial improvement and give state-of-the-art results on benchmark data sets such as GLUE, SQuAD, and RACE data-sets.

%Recently, Devlin et al. \cite{devlin2018bert} introduced the Bidirectional Encoder Representation from Transformer (BERT) that leverages Transformer based architecture. BERT has been pre-trained on English Wikipedia (2,500 million words), it can be fine-tuned to solve downstream NLP tasks, and this process of fine-tuning is also known as Transfer Learning. BERT has been successfully applied on various NLP tasks such as Sentiment Analysis, Language Understanding, and Natural Language Inference. BERT considers the full context of a word by looking at the words that come before and after it. Despite the achievements of the BERT model in NLP tasks, operating BERT under constrained computational power can be challenging. To reduce the computational power required to train BERT Sanh et al. \cite{sanh2019distilbert} introduced a smaller pre-trained version of BERT called DistilBERT. It has 40\% fewer parameters and is 60\% faster than the BERT while retaining 97\% of the accuracy. DistilBERT leverages knowledge distillation during the pre-training phase which enables to approximate BERT while retaining the almost same accuracy.
\section{Methodology}
\label{sec:methodology}

%please note this is first draft of the methodology section, errors expected!

\begin{figure}[htbp]
\centerline{\includegraphics[width=0.5\textwidth]{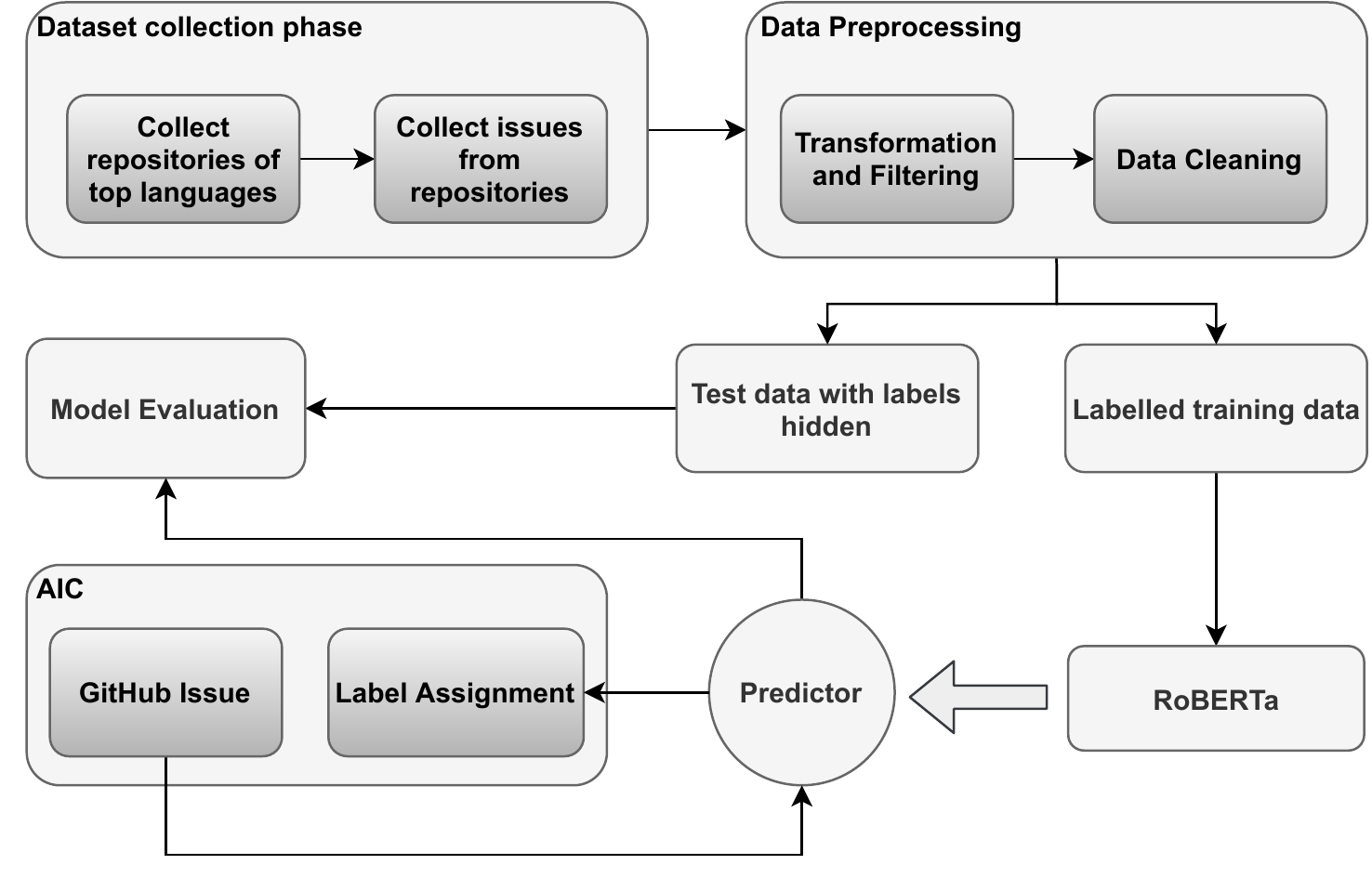}}
\caption{Overview of the Framework: Starts with data collection phase. Further, gathered dataset is pre-processed. The pre-processed data is then splited into train and test sets in order to evaluate the results.}
\label{fig:overview}
\end{figure}

This section presents our proposed approach towards the classification of issue reports on GitHub. Our approach consists of the following phases: (1) Data Collection phase, which involves data collection from GitHub (2) Data Pre-processing phase, which involves data transformation and cleaning tasks. (3) Model Construction phase, which involves training a multi-label transformer-based model i.e. RoBERTa. (3) Automatic Issue Classifier (AIC), a GitHub application which can be leveraged to classify GitHub issues-reports on any GitHub repository. Figure \ref{fig:overview} shows the overview of our approach. We briefly highlight the different phases of our approach in this section and explain details of each phase in the subsequent subsections.

%This section presents the approach towards the automated solution for classifying issues on GitHub. First, we discuss the methods used for the construction of the data set. This comprises collecting, cleaning, filtering, and transforming the data set in order to prepare it for the classification task. The second part focuses on building the model framework where we use state-of-the-art, attention-based natural language classification technique; RoBERTa and fine-tune it on textual data from GitHub issues. Finally, we present a tool that automatically labels any newly reported issues once added to any GitHub repository.

\subsection{Data Collection}
For this work, we used GitHub as our data source. First, we identified the top 55 candidate popular languages using the IEEE Spectrum Top Programming Languages list \cite{languageranking}. Further, we fetched the top 200 popular repositories across each programming language. Each repository on GitHub has an associated star count, which represents the popularity of the repository \cite{repopopularity}. Here, we identified the project by their primary language. The primary language of the project is defined as the one in which the largest number of source code files are written \cite{8870174}. Lastly, we fetched issues associated with each repository along with the related attributes. These attributes include, but are not limited to; title, body, and labels of issues.

\begin{table}[]
\centering
\begin{tabular}{|c|c|}
\hline
\textbf{Language} & \textbf{Projects} \\ \hline
Assembly & Apollo-11, MS-DOS, mal \\ \hline
C & linux, netdata, redis \\ \hline
C++ & tensorflow, electron, terminal \\ \hline
PHP & laravel, jQuery-File-Upload, SecLists \\ \hline
JavaScript & freeCodeCamp, vue, javascript \\ \hline
Go & go, kubernetes, awesome-go \\ \hline
Ruby & rails, jekyll, discourse \\ \hline
Dart & flutter, awesome-flutter, flutter-go \\ \hline
C\# & shadowsocks-windows, PowerToys, PowerShell \\ \hline
Racket & pollen, hackett, frog \\ \hline
%Lua & kong, waifu2x, neural-style \\ \hline
Objective-C & AFNetworking, SDWebImage, GPUImage \\ \hline
%Fortran & OpenBLAS, pymc, lapack \\ \hline
Scala & spark, prisma1, scala \\ \hline
\end{tabular}
\caption{Top 3 GitHub repositories associated across different languages.}
\label{top_project}
\end{table}

Our data set contains issues from numerous open-source and industrial projects such as Laravel, Kubernetes, Linux, Spark. Table \ref{top_project} shows the top three projects from our data across some of the programming languages. In total, we were able to fetch 1,166,107 issues out of which 509,090 of the issues were unlabelled and 657,017 issues were labeled. The substantial amount of unlabelled issues provides us the opportunity to further dig down to this problem and provide an effective approach to categorize the issue reports. We used GitHub Search and Data API \footnote{\url{https://docs.github.com/en/rest}} to fetch this dataset. The complete dataset is available for public use \footnote{\url{https://www.kaggle.com/ansnadeem/github-issues}}.

\subsection{Data Pre-processing and Transformation}

\textit{Transformation and Filtering :} We extracted the title and body of the issue and concatenated them together. We filtered our GitHub issue reports into three categories i.e. bug-report, enhancement, and question consistent with the prior work by Kallis et al.\cite{KALLIS2021102598}. Here, each label is non-exclusive such that an issue can belong to more than one category. For instance, an issue can be labeled as a bug report and enhancement at the same time. Table \ref{multilabeledexample} shows one such issue example.

\begin{table}
\centering
\resizebox{\linewidth}{!}{%
\begin{tabular}{|>{\hspace{0pt}}m{0.344\linewidth}|>{\hspace{0pt}}m{0.374\linewidth}|>{\hspace{0pt}}m{0.211\linewidth}|}
\hline
\textbf{Title} & \textbf{Body} & \textbf{Categories} \\
\hline
USBhost: additional functions for keyboard~appreciated & for USBhost it would\par{}be fine to have additional\par{}functions to read the \par{}USB keyboard:kbhit()\par{} getch() getche()\par{}getchar() gets() scanf() & bug,\par{}enhancement \\
\hline
\end{tabular}
}
\caption{Issue report that is labelled with both bug and enhancement categories.}
\label{multilabeledexample}
\end{table}

The labels under consideration in our research are the default GitHub labels:
\begin{enumerate}
    \item \textbf{Bug-report:} Issue that reports a bug or fault in the project.
    \item \textbf{Enhancement:} Issue that requests the enhancement of the project, such as performance improvement, feature request.
    \item \textbf{Question:} Issue that asks the question from the repository owner or maintainer.
\end{enumerate}

We further filtered out the issues that were not in the English language and only used issues that are written in English. We used langdetect \footnote{\url{https://github.com/Mimino666/langdetect}} to tag every issue with the language used in the text and found that 647,438 of the labeled issues were written in English while only 9,579 issues were written in other languages. %Finally, to incorporate samples from diverse project types, we randomly sampled 100,000 issues. %Our resultant data-set contains 120081 bugs, 100562 enhancements, and 21451 questions.

\textit{Pre-processing:} We did not perform traditional NLP pre-processing steps such as stop-word removal, stemming, and lemmatization. The RoBERTa has its own tokenizer that does not require traditional NLP pre-processing steps to be applied to the text. However, we converted the text of all issue reports to lower case, as we are using RoBERTa with its base pre-trained model which is a case-sensitive model. Additionally, as our data set contains real-world issue reports, our data set has a class-imbalance problem. Less than 4\% issues were labeled as questions. To resolve, the class imbalance problem, we randomly over-sampled the issue reports with less representative labels i.e. question label.

\subsection{Model Construction}

Pre-trained transformer-based models are trained on a large natural language corpus, which empowers them to extract the detailed contextual word representation from the text. We used one such transformer-based model called RoBERTa \cite{roberta}, a robust and optimized variant of BERT \cite{devlin2018bert}. The reason behind choosing RoBERTa over BERT is its state-of-the-art results on benchmark data sets such as GLUE, RACE, and SQuAD with over 2-20\% improvement over BERT. RoBERTa has several configurations which can be tuned on the basis of the pretraining approach. We use `roberta-base' which consists of 12 transform layers, 768 hidden layers, and 12 self-attention heads. We used simpletransformers \footnote{https://github.com/ThilinaRajapakse/simpletransformers/} to fine-tune our model. The package uses HuggingFace's \footnote{\url{https://huggingface.co/}} implementation of the model.

\textit{Model Training:} As we are using a pre-trained model, which essentially requires less number of training examples to be fine-tuned as compared to model trained from scratch, we randomly sampled 55,000 (approx) issue reports. We then divided the issue data set into 80\%-20\% train-test split and fed the training data set to fine-tune our model. We trained our model for 5 epochs with a learning rate of 4e-05, a maximum sequence length of 128, and a batch size of 8. We trained our model using Google Colab \footnote{\url{https://colab.research.google.com/}}. With 12 GB RAM, Tesla T4 14 GB GPU, and 2 Intel Xeon CPUs @ 2.2GHz it took an hour to train the model.

%Figure \ref{fig:BERT} shows the overview of the BERT fine-tuning architecture for our multi-label classification.

% Attention-based transformers are usually pre-trained on text such as a large English corpus or textual data collected from Wikipedia, and then fine-tuned for a specific task. This gives the attention mechanism ability to be context-aware and hence determine the important details in an input sequence corresponding to the context. We use RoBERTa\cite{roberta} which is an optimized and robust variant of BERT. We use this variant because it achieves state-of-the-art results on benchmarks such as GLUE, RACE, and SQuAD with over 2-20\% improvement atop BERT. RoBERTa has several configurations which are differentiated on the basis of the pretraining approach. We use RoBERTa base which consists of 12 layers, 768 hidden layers, and 12 transformer heads. We use simpletransformers to construct our Multi-Label Classification model framework. Due to the pre-trained nature of our model, we first shuffle the dataset and further slash our working data-set by keeping only 15000 issues associated with every label. This also gets rid of the high-class imbalance. We then split the obtained data-set with a threshold of 0.8 and 0.2 and train it with 10 epochs. We observe state-of-the-art results even without including all the training examples, with the requirement of a short training time.

\subsection {Automatic Issue Classifier}
\begin{figure}[htbp]
\centerline{\includegraphics[width=0.4\textwidth]{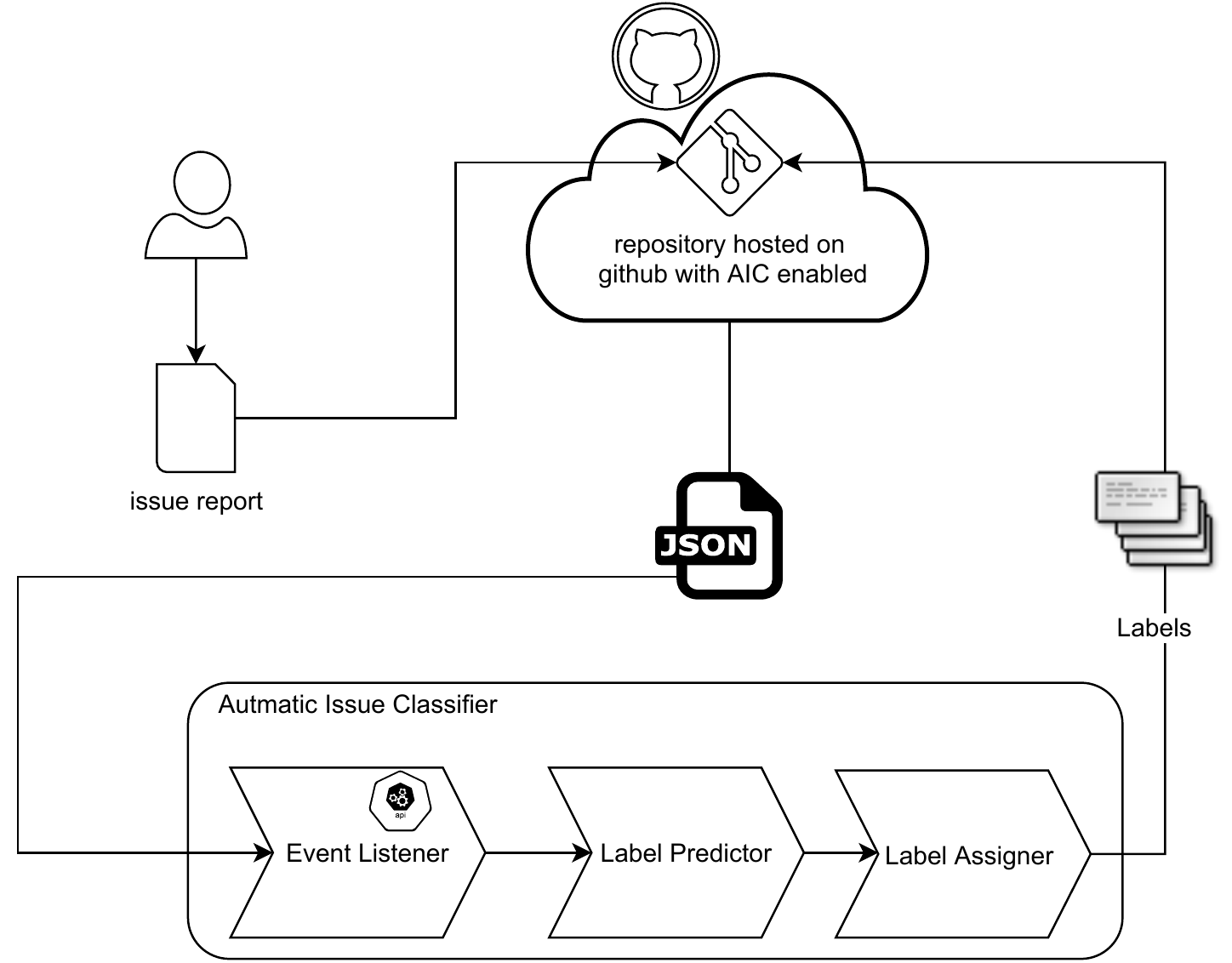}}
\caption{AIC sits idle until a user creates an issue on a repository integrated with AIC. A new issue sends an event to AIC API, which in results predicts the issues based on the text in the issue report and assigns the label to it.}
\label{fig:aic_flow}
\end{figure}

We deployed our model using a GitHub application called Automatic Issue Classifier (AIC). Figure \ref{fig:aic_flow} explains the working flow of the application. AIC application can be integrated into any GitHub repository and can automatically assign the newly created issue report with its respective label. AIC is a Flask \footnote{\url{https://flask.palletsprojects.com/en/2.0.x/}} web framework based application with a python run-time and is deployed on a WSGI server \footnote{\url{https://wsgi.readthedocs.io/en/latest/index.html}}. AIC application can be integrated with any existing GitHub repository by accessing it through the application page \footnote{\url{https://github.com/apps/automatic-issue-classifier}}. After landing on the application page, click on the 'Install' button and select the GitHub repository you want to associate it with. %Figure \ref{issue_label_demo} shows the Issue Classifier application in work after integration with a GitHub repository.

%Finally, we develop a GitHub application, based on our research. Our application a tool that can be integrated with any GitHub repository. It is a Flask application with a python run-time and is deployed on a WSGI server. Our tool automatically classifies any newly reported issues and assigns it appropriate labels. The application can be integrated with any existing GitHub repository by accessing it through the application page \footnote{\url{APPLICATION_LINK_HERE}}, clicking 'Install' and finally selecting the repositories to integrate it with.

%Any newly reported issue on a repository generates events for our application. In response, our application fetches the issue-text and utilizes it to predict corresponding labels and then sends the predicted labels back to GitHub in order to assign them to the newly created issue.
\section{Evaluation}
\label{sec:evaluation}

In this section, we evaluate our proposed approach using GitHub issue reports data set and answer the research questions presented earlier in section \ref{sec:introduction}. Following metrics are used to evaluate the proposed approach:.
%In addition to accuracy highlighted in the Figure \ref{accuracy_compare} for each of the models trained on our data set, we also use the test data to evaluate the following metrics for our models:

%\begin{figure}[htbp]
%\centerline{\includegraphics[width=0.5\textwidth]{Figures/bar_comparison.png}}
%\caption{Accuracy comparison over results of all models with each of the three labels}
%\label{accuracy_compare}
%\end{figure}

\begin{itemize}
    \item \textbf{Precision} is the ratio of true positives to the sum of true positives and false positives. A precision value closer to 1 is the most desirable.
    %\[Precision = \frac{TP}{TP+FP}\]

    \item \textbf{Recall} is a measure of how well we were able to identify relevant instances. It is defined as the ratio of true positives to the sum of true positives and false positives. A recall value of 1 indicates the best performance.
    %\[Recall = \frac{TP}{TP+FN}\]

    \item \textbf{F1-Score} is defined as the harmonic mean of precision and recall. F-1 score closer to 1 is the most preferable.
    %\[F1 = 2 * \frac{precision*recall}{precision+recall}\]

    \item \textbf{Hamming Loss} is a commonly reported metric for multi-label classification. It is the fraction of incorrectly predicted labeled out of the total number of labels. A hamming loss value of 0 indicates the best results.
    %\[HL = \frac{1}{N L} \sum_{l=1}^L\sum_{i=1}^N Y_{i,l} \oplus X_{i,l}\]

\end{itemize}

\textbf{RQ1: Does the transformer-based approach perform better than the traditional keyword-based approach at classifying GitHub issue reports?}

Table \ref{resultstable} shows the results of our approach over GitHub issue reports classification. Our RoBERTa based approach is able to predict results for bugs, enhancement, and questions category with a hamming loss of 0.17, 0.16, and 0.16 respectively. Similarly, we were able to achieve an F1 score of 81\%, 74\%, and 80\% for bugs, enhancement, and questions category respectively which are comparable to results of existing studies. For instance, Kallis et al.\cite{KALLIS2021102598} reported F1 scores of 83.1\%, 82.3\%, 82.5\% for bugs, enhancement, and questions category respectively. It is observed that our results are slightly degraded than the results achieved by Kallis et al.\cite{KALLIS2021102598}. This is expected because our framework proposes a multi-label approach as opposed to the multi-class approach. Moreover, they sampled data in a class if any assigned label contains the text 'question', 'bug' or 'enhancement' however label text can be 'not-a-bug' which in this case would also be incorrectly sampled as a bug report, our sampling, in contrast, is based on exactly matching these labels resulting in more accurate classification. %Secondly, a lesser number of samples are associated with the questions category in the original dataset, categorized as questions. Only 3,682 of the 100,000 issues we sampled were labeled as questions which are less than 4\%. We believe the reason for this data imbalance in the question category is primarily due to the common use of other platforms such as StackOverflow \cite{github_stackoverflow} at a higher frequency, to ask questions.

\begin{table}[]
\centering
\begin{tabular}{|c|c|c|c|c|}
\hline
\textbf{Category} & \textbf{Precision} & \textbf{Recall} & \textbf{F1-Score} & \textbf{Hamming Loss} \\ \hline
Bug & 81\% & 81\% & 81\% & 0.14 \\
Enhancement & 78\% & 72\% & 74\% & 0.15 \\
Question & 79\% & 81\% & 80\% & 0.15 \\
Macro-Average & 79\% & 78\% & 78\% & 0.15 \\ \hline
\end{tabular}
\caption{Evaluation results of our proposed approach}
\label{resultstable}
\end{table}

%The tables \ref{eval_nb} for NaiveBayes-baseline, \ref{eval_log} for LogsiticRegression and \ref{eval_svc} for LinearSVC show the evaluation metrics computed over the test data. Our models have high accuracy as indicated in figure \ref{accuracy_compare} however, we were able to achieve 0.52, 0.63, and 0.48 for LinearSVC of feature, bug, and enhancement labels. We believe that the main reason for this is an imbalance of classes in the training examples. We also observe that our LinearSVC performs better than our baseline model as well as logistic regression-based model.

%\input{Tables/naive_bias_results}

%\input{Tables/lr_results}

%\input{Tables/svc_results}

\textbf{RQ2: Does the transformer-based model have the ability to be incorporated into an industry-level tool?}

\begin{figure}[htbp]
\centerline{\includegraphics[width=0.5\textwidth]{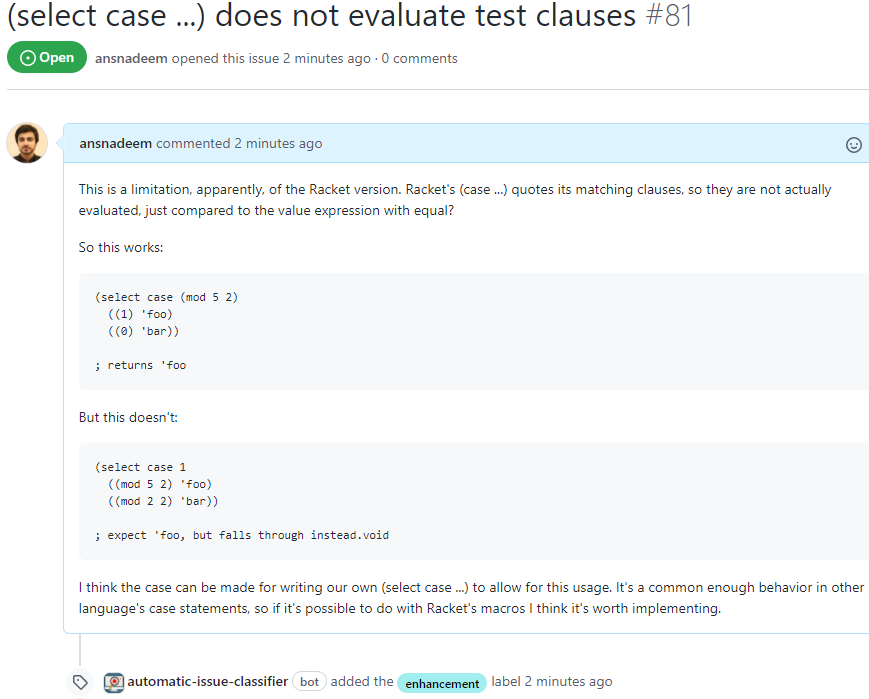}}
\caption{Automatic Issue Classifier automatically assigns label to a newly reported issue on a repository.}
\label{fig:issue_label_demo}
\end{figure}

We successfully adopted and deployed the transformer-based model in an industry tool called Automatic Issue Classifier (AIC). Figure \ref{fig:issue_label_demo} shows a demonstration of AIC automatically assigning a label to an issue report. When a user creates an issue, AIC automatically assigns it a label based on the context. We deployed the tool as a GitHub application to ensure easy integration with industry projects. To enable users to tailor the application according to specific requirements, the complete source code of our application is available via a GitHub repository\footnote{\url{https://github.com/ansnadeem/aic}}.

\section{Relevance to the Industry}
\label{sec:discussion}
%In this section, we discuss the results of our approach. While evaluating, we treat eac

%This section discusses the use of these results in the industry.

Effective classification of issue reports can have an immediate effect on software development processes. First, it can enable easy tracking of issue reports on GitHub, which could result in effective software maintenance. Secondly, the volume of each categorized issue report can help the project manager to review the project maintenance process at any point and make appropriate decisions. For instance, a large number of bug reports might indicate that the quality assurance process might need some help. Thirdly, having accurately categorized issue reports would help in routing the issue to the right software developer, who is working on the specific module/component. Also, accurately categorized issue reports can help the team leader to prioritize the tasks, and the release manager can figure out the way to stabilize the project. Lastly, the projects will be managed more efficiently and effectively, and ultimately would result in better products and projects.

%Following are some of the reports that can be generated:
%\begin{enumerate}
%    \item \textbf{Developer Report:} The developer reports could include the issue reports assigned to each developer. This will give the project manager an idea who is responsible of resolving the specific issue-report.
%    \item \textbf{Project Report:} The project report can show the volume of each categorized issue-reports along with the source-code associated with it. The goal is to provide holistic real-time progress of the project maintenance to managers. 
%\end{enumerate}

%\input{relevance_to_the_industry}
\section{Threats to Validity}
\label{sec:threats}
This section highlights some threats in validating our work.
\begin{itemize}
    \item \textbf{Popularity Bias:} We are fetching results from the top repositories of the popular programming languages, so our results might be influenced by the popularity bias.
    \item \textbf{Language Imbalance Bias:} Our data set implies that some languages on GitHub are more popular than the rest. For instance, our data set contains a significant portion of the issues-reports that are associated with C\# repositories.
\end{itemize}

\section{Future Work}
\label{sec:future_work}
Our future work would revolve around improving our industry tool: 
\begin{itemize}
    %\item \textbf{Include More Labels}
    %Right now our framework predicts three GitHub labels which are bug, question and enhancement. In future, we intend to add more labels for example ‘documentation’ and analyze if our existing model can predict those labels

    \item \textbf{Automatically Assigning Issues}
     We plan to extend our tool by making it capable of not only assigning labels but also assigning them to a developer.
    
%    \item \textbf{Analyzing Misclassified Issues:} We plan to conduct a detailed study on the miscategorized issues across different languages to identify and address the root cause of misclassification.
        
    \item \textbf{More Categories:} We also plan to add more categories such as `documentation', `wontfix', and `help wanted', `duplicate' in the issue data set, and analyze if our model can also categorize them with similar accuracy.
    
    %\item \textbf{Dashboard over AIC:} We plan to incorporate AIC in a dashboard that will allow the software team to review the software development and software maintenance process.

\end{itemize}

\section{Conclusion}
\label{sec:conclusion}
We proposed a transformer-based approach towards the classification of GitHub issues and assigning them respective labels. Our model is able to well understand the context, hence provides accurate and efficient issue report classification. Our framework takes into account the non-exclusiveness of the issue-reports data and is capable of assigning multiple labels to issue reports at the same time. We extracted our data-set from GitHub repositories across the top 55 popular programming languages to ensure that our results and model are capable of generalizing across multiple languages. We evaluated our results over a data set containing three labels i.e., bugs, enhancement, and questions, and were able to achieve an F-1 score of 81\%, 74\%, and 80\% respectively, which is comparable to existing issue classification results. We leveraged our state-of-the-art approach to develop an industry tool i.e. Automatic Issue Classifier that is capable of automatically labelling the issue reports. We believe that our tool will benefit the software engineering industry by enabling the teams to keep their focus on the necessary tasks by automating the manual task of labelling the issue reports.

%which help to track, prioritize and execute the software maintenance activities.

\section{Acknowledgements}
\label{sec:conclusion}
This work was supported by North Dakota State University College of Engineering. We would also like to thank Giorgos Karantonis\footnote{\url{https://github.com/GiorgosKarantonis/}} for sharing his key insights regarding the research problem.

%We also acknowledge the work of Karantonis, G. on Github Issues Classifier\footnote{\url{https://github.com/GiorgosKarantonis/Github-Issues-Classifier}} which employed a similar approach for a project.

\bibliographystyle{IEEEtran}
\bibliography{bibliography}

\end{document}